\documentclass[%
 reprint,
superscriptaddress,
 amsmath,amssymb,
 aps,
prb,
]{revtex4-2}
\usepackage{lineno}
\usepackage[utf8]{inputenc}
\usepackage{textcomp}
\usepackage{subfigure}
\usepackage{graphicx}
\usepackage{epstopdf}
\usepackage{dcolumn}
\usepackage{bm}
\usepackage{gensymb}
\usepackage{upgreek}
\usepackage{hyperref}
\usepackage{tabularx}

\usepackage{xcolor}
\usepackage{xparse,xcoffins}
\usepackage{verbatim}

\ExplSyntaxOn
\NewCoffin\imagecoffin
\NewCoffin\labelcoffin

\ExplSyntaxOff

\preprint{APS/123-QED}

\begin{document}

\title{Tuning Incommensurate Charge Order in Ba$_{1-x}$Sr$_x$Al$_4$ and Ba$_{1-y}$Eu$_y$Al$_4$}

\author{Prathum Saraf}
\affiliation{Maryland Quantum Materials Center, Department of Physics, University of Maryland, College Park, Maryland 20742, USA}
\author{Eleanor M. Clements}
\affiliation{NIST Center for Neutron Research, National Institute of Standards and Technology, Gaithersburg, Maryland 20899, USA}
\author{Danila Sokratov}
\author{Shanta Saha}
\affiliation{Maryland Quantum Materials Center, Department of Physics, University of Maryland, College Park, Maryland 20742, USA}
\author{Peter Zavalij}
\affiliation{Department of Chemistry, University of Maryland, College Park, Maryland 20742, USA}
\author{Thomas W. Heitmann}
\affiliation{University of Missouri Research Reactor, University of Missouri, Columbia, Missouri 65211, USA}
\author{Jeffrey W. Lynn}
\affiliation{NIST Center for Neutron Research, National Institute of Standards and Technology, Gaithersburg, Maryland 20899, USA}

\author{Camille Bernal-Choban}
\author{Dipanjan Chaudhuri}
\author{Caitlin Kengle}
\author{Yue Su}
\author{Simon Bettler}
\author{Nathan Manning}
\author{Peter Abbamonte}
\affiliation{Department of Physics and Materials Research Laboratory, The Grainger College of Engineering, University of Illinois Urbana–Champaign, Urbana, 61801, IL, USA}

\author{Sananda Biswas}
\affiliation{Institut f\"ur Theoretische Physik, Goethe-Universit\"at Frankfurt, 60438 Frankfurt am Main, Germany}
\author{Roser Valent\'\i}
\affiliation{Institut f\"ur Theoretische Physik, Goethe-Universit\"at Frankfurt, 60438 Frankfurt am Main, Germany}
\affiliation{Canadian Institute for Advanced Research, Toronto, Ontario M5G 1Z8, Canada}
\author{Johnpierre Paglione}
\affiliation{Maryland Quantum Materials Center, Department of Physics, University of Maryland, College Park, Maryland 20742, USA}
\affiliation{Canadian Institute for Advanced Research, Toronto, Ontario M5G 1Z8, Canada}

\date{\today}

\begin{abstract}

The BaAl$_4$-type structure family is home to a vast landscape of interesting and exotic properties, with descendant crystal structures hosting a variety of electronic ground states including magnetic, superconducting and strongly correlated electron phenomena.  
BaAl$_4$ itself hosts a non-trivial topological band structure, but is otherwise a paramagnetic metal. However, the other members of the $A$Al$_4$ family ($A$= alkali earth), including SrAl$_4$ and EuAl$_4$, exhibit symmetry-breaking ground states including charge density wave (CDW) and magnetic orders.
Here we investigate the properties of the solid solution series Ba$_{1-x}$Sr$_x$Al$_4$ and Ba$_{1-y}$Eu$_y$Al$_4$ using transport, thermodynamic and scattering experiments to study the evolution of the charge-ordered state as it is suppressed with Ba substitution to zero near 50\% substitution in both systems. 
Neutron and x-ray diffraction measurements reveal an incommensurate CDW state in SrAl$_4$ with $c$-axis-oriented ordering vector (0, 0, 0.097) that evolves with Ba substitution toward a shorter wavelength. A similar progression is observed in the Ba$_{1-y}$Eu$_y$Al$_4$ series that also scales with the ordering temperature, revealing a universal correlation between charge-order transition temperature and ordering vector that points to a critical wavevector that stabilizes CDW order in both systems.
We study the evolution of the phonon band structure in the Ba$_{1-x}$Sr$_x$Al$_4$ system, revealing the suppression of the CDW phase matches the suppression of a phonon instability at precisely the same momentum as observed in experiments, confirming the electron-phonon origin of charge order in this system. 

\end{abstract}

\maketitle

\section{Introduction}

The evolution of interest in systems harboring charge-ordered states has produced a plethora of fascinating  phenomena and observations that shed light on both cooperative and competing states in a wide range of materials.
Starting with the discovery and understanding of the role of charge order in the cuprate phase diagram \cite{Comin2016}, and more recently with the varieties of charge order in transition metal dichalcogenides \cite{Lin2020}, time reversal symmetry breaking at the charge order transition in the Kagome series $A$V$_3$Sb$_5$ \cite{Wilson2024},
interplay with strong correlations in FeGe \cite{Teng2022} and nematic order and fluctuations in BaNi$_2$As$_2$ \cite{Eckberg-Nematic}, the interplay of charge, spin and orbital degrees of freedom has come into focus as a key question in quantum materials.
Additionally, the effect of charge modulation and symmetry breaking on topological aspects of electronic band structures has come to light as a potential area of interest, for instance in kagome materials \cite{Neupert2021}, and charge-ordered phases have themselves been proposed to entail non-trivial topologies \cite{TiSe2-topology,litskevich2024discovery}.

\begin{figure}
{\includegraphics[scale=0.4,width=0.45\textwidth]{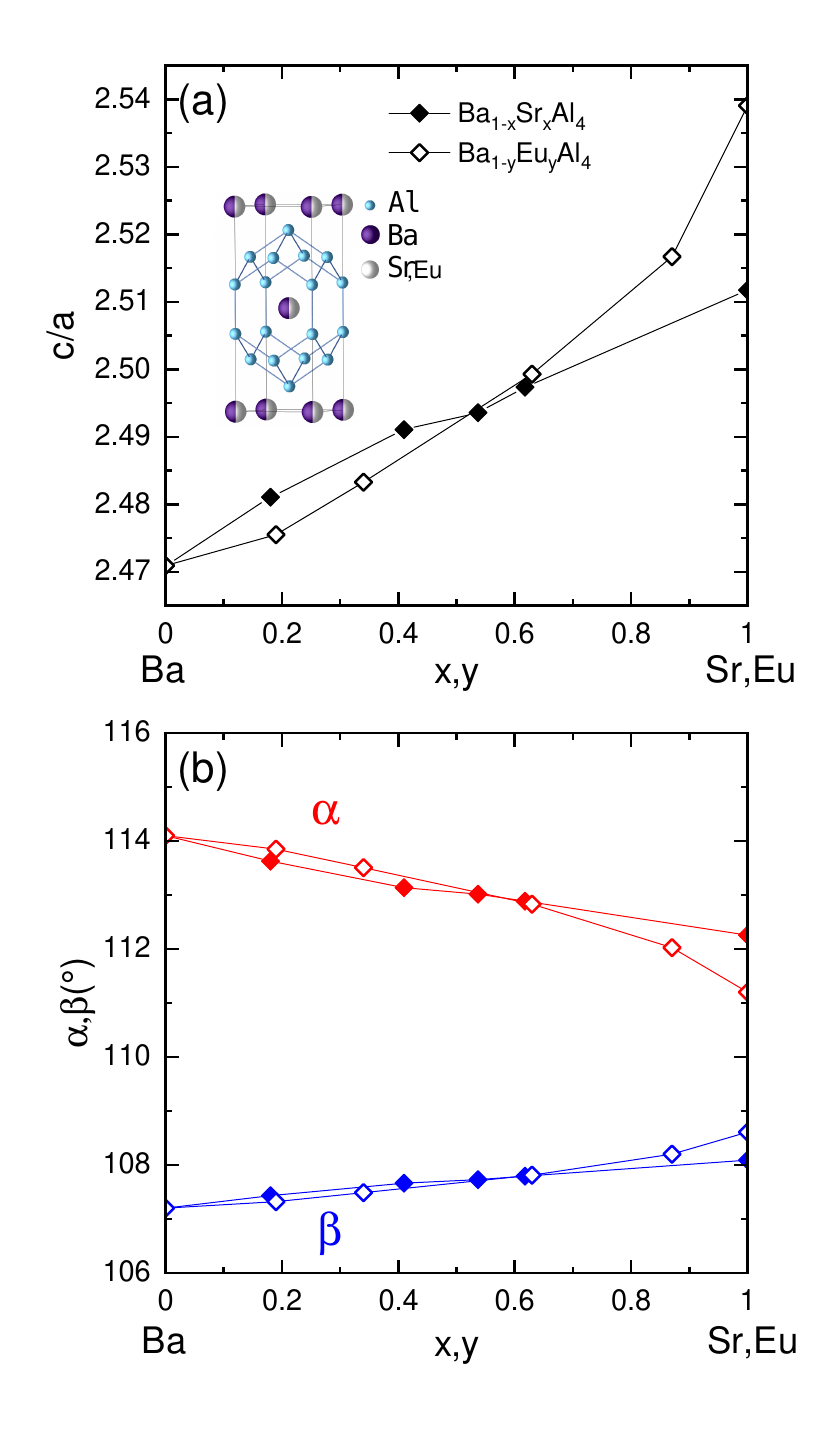}}
\caption{Plot of quantities measured through X-ray experiments. (a) Plot of the ratio of c-axis to a-axis. (b) The blue and red curves are plotting the angles $\alpha$ and $\beta$, respectively, which are the bond angles in the aluminum tetrahedron.}
\label{fig: X-ray}
\end{figure}

The family of materials with the BaAl$_4$ parent structure type hosts a variety of exotic behaviors \cite{BaAl4structures}, including iron-based superconductivity \cite{PaglioneFeSCs}, heavy-fermion physics in systems such as CeCu$_2$Si$_2$ \cite{CeCu2Si2}, electronic nematic phases in BaNi$_2$As$_2$ \cite{Eckberg-Nematic}, hidden order in URu$_2$Si$_2$ \cite{Palstra-URS,Maple-URS,Schlabitz-URS}, and topological spin textures in EuGa$_2$Al$_2$ \cite{Moya-EuGa2Al2} and EuCd$_2$As$_2$ \cite{Valadkhani}. 
The $A$Al$_4$ and $A$Ga$_4$ series of binary compounds - with $A$= Ba, Sr, Ca or Eu - have been known to harbor a variety of charge-ordered and magnetic ground states 
\cite{NAKAMURA-SrAl4,Nakamure-EuGa4}, including charge density wave (CDW) order at 243~K and 140~K in SrAl$_4$ and EuAl$_4$, respectively, and rare earth magnetism below $\sim 20$~K in the Eu-based compounds.
However, recent studies of the BaAl$_4$ compound itself -- a non-magnetic, metallic material that does not exhibit any charge-ordered phase -- have revealed it to host a 
crystalline symmetry-protected non-trivial topology with three-dimensional Dirac spectrum \cite{BaAl4-Topology,BaAl4-Topology2} that likely also exists in the Sr and Eu-based counterparts \cite{Maia-Topological}.
Charge order in SrAl$_4$ and EuAl$_4$, as well as its  anomalous absence in the nearly identical BaAl$_4$, have recently been studied theoretically \cite{Wang2024} and experimentally \cite{Kaneko2021,Ramakrishnan-EuAl4,Korshunov,Ramakrishnan,Yang2024}, and debate about the origins of CDW order suggest a subtle sensitivity of the CDW phase to details of the phonon and electron band structures \cite{Wang2024,Ramakrishnan}.

In this work, we study the evolution of the CDW state in the Ba$_{1-x}$Sr$_x$Al$_4$ and Ba$_{1-y}$Eu$_y$Al$_4$ solid solution series as a function of isovalent Sr/Eu substitution ($x$) in order to verify and fully characterize the evolution of charge order with structural and electronic properties between the disparate end-members, and to identify how CDW order appears with Sr and Eu substitution. Using single-crystal x-ray and neutron diffraction experiments, we characterize the CDW ordering $q$-vector as a function of temperature and $x$ for several characteristic concentrations  through the range where CDW order is present in each series, to observe an evolution of the incommensurate $q$-vector toward the complete suppression of CDW order near $x$=0.50 in each case. We also performed x-ray diffraction experiments and $ab~initio$ density functional theory (DFT) calculations on the non-magnetic Ba$_{1-x}$Sr$_x$Al$_4$ solid solution series to show how the minimal change in crystal structure with alkali earth substitution has an impact on the phonon spectrum, resulting in an acoustic mode instability that is suppressed in parallel with the observed CDW order.
Our results are in good agreement with previous work suggesting CDW order in this series originates from a softened phonon mode and strong electron-phonon coupling, and provide detailed insight into how CDW order is universally fine-tuned as a function of divalent cation substitution.  

\section{Methods}

Single crystals of Ba$_{1-x}$Sr$_x$Al$_4$ and Ba$_{1-y}$Eu$_y$Al$_4$ were grown using an Al self-flux method from elemental Ba, Eu, Sr and Al (purities: 99\%, 99.9\%, 99\% and 99.999\% respectively) in a ratio of 1-$x$:$x$:16, with excess Al to act as a flux at high temperatures. The combination was placed in an alumina crucible and sealed inside a quartz tube in an argon atmosphere. The reaction mixture was heated to 1100 $^{\circ}$C at 50 $^{\circ}$C/hour, held at 1100 $^{\circ}$C for 12 hours to homogenize the mixture and slow cooled to 720 $^{\circ}$C at 5 $^{\circ}$C/hr before being put in a centrifuge to separate the flux from the crystals. This process resulted in large single crystals with dimensions exceeding 2~$\times$~2~$\times$~1~mm$^3$, with the shortest axis universally being the \textit{c}-axis, typical for tetragonal systems.
Electrical transport measurements were made with a Quantum Design Physical Properties Measurement System (PPMS). For the transport measurements, gold wires were attached to the samples with DuPont 4929 silver paste.

Single-crystal X-ray diffraction at 250~K was collected on  Ba$_{1-x}$Sr$_x$Al$_4$ and Ba$_{1-y}$Eu$_y$Al$_4$ samples with a Bruker APEX-II CCD system equipped with a graphite monochromator and a MoK$\alpha$ sealed tube ($\lambda$ = 0.71073 $\textup{\AA}$). One sample ( SrAl$_4$, UM$\#$4008) was measured using a Bruker D8 Venture Duo diffractometer equipped with Photon III detector, MoK$\alpha$ micro-focus tube and Helios optics, and all crystal structure refinement was performed using the Bruker ShelXTL software package.
Temperature-dependent single-crystal x-ray diffraction measurements on Ba$_{1-y}$Eu$_y$Al$_4$ using a three-dimensional (3D) Mo K$\alpha$ (17.4-KeV) source which delivers on the order of 10$^7$ photons per second with a beam spot of 130 $\mu$m. The sample was cooled using a closed-cycle cryostat to reach a base temperature of 12 K. The sample was kept inside a Be dome, which was used for vacuum and radiation shielding. Sample motion was performed using a Huber four-circle diffractometer and X-ray detection was captured using a Mar345 image plate to allow for 3D mapping of momentum space with a resolution of q = 0.01 to 0.08 $\textup{\AA}^{-1}$ depending on the cut. We transformed the data collected from real to reciprocal space over volumes varying from V = 0.0007 - 0.003 $\textup{\AA}^{-3}$, depending on the zone of the Bragg peak. These volumes encompass the entire intensity of the CDW peak at base temperature, and were kept constant (with increasing temperatures) for each substitution concentration.

Neutron diffraction measurements were performed with the BT-7 triple axis spectrometer at the NIST Center for Neutron Research using a fixed incident wavelength $\lambda$ = 2.359 $\textup{\AA}$, horizontal collimations set to open -50'-50'-120' (full width at half maximum), pyrolytic graphite monochromator and analyzer, and pyrolytic graphite filters to suppress higher order wavelengths \cite{Lynn-Neutron}.  A 57 mg single crystal of Sr$_x$Al$_4$ was aligned in the tetragonal (space group I4/mmm) [H,0,L] scattering plane, where $a^*$ = 2$\pi$/a and $c^*$ = 2$\pi$/c with a = 4.6424 $\textup{\AA}$ and c = 11.2148 $\textup{\AA}$ at 300 K. To reduce the possibility of introducing strain to the sample, we avoided using wire or glue for mounting. Instead, the crystal was mounted to an aluminum post by gently wrapping it with aluminum foil. A closed-cycle refrigerator with a base temperature of 5 K was used to control the sample temperature. Scans in reciprocal space were measured as a function of H and L on warming from 5 – 255 K.  A very similar configuration was employed at the University of Missouri Research Reactor, with collimations of 60'-60'-40'-40'.  Uncertainties, where indicated throughout, represent one standard deviation.

\section{Experimental Results}

\begin{figure*}[t]
\hfill
\includegraphics[scale=0.1,width=0.98\linewidth]{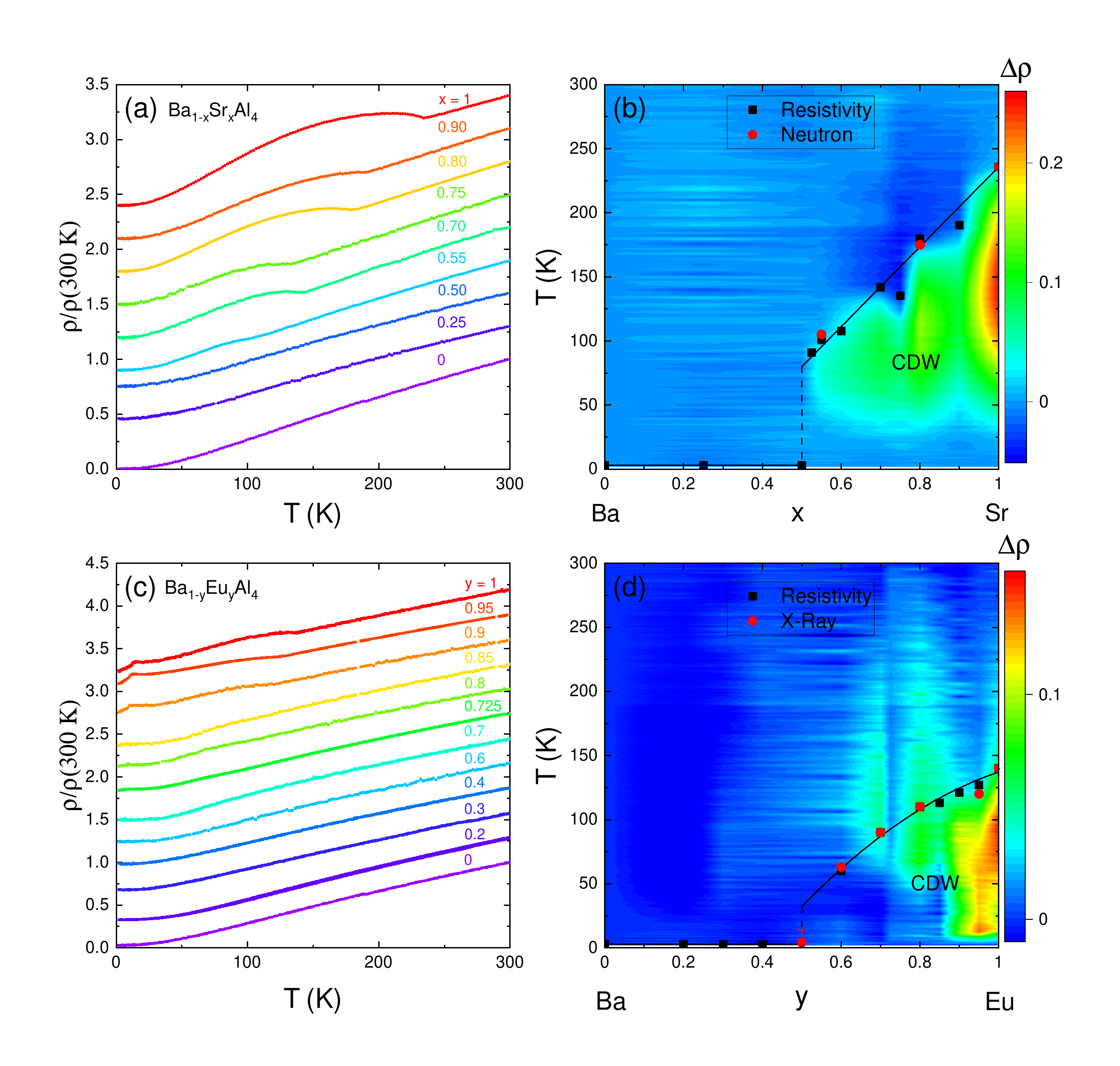}
\hfill
   \caption{(a) and (c) Resistance curves of various substitutions of Ba$_{1-x}$Sr$_x$Al$_4$ and Ba$_{1-y}$Eu$_y$Al$_4$ respectively. The curves are normalized and stacked for easier interpretation. Suppression of the CDW transition is seen with hints of it in x = 0.55 substitution before disappearing at x = 0.5 in Ba$_{1-x}$Sr$_x$Al$_4$ and a similar suppression is seen around y = 0.4 in Ba$_{1-y}$Eu$_y$Al$_4$. Panels (b) and (d) present phase diagrams of both substitution series, constructed using the resistivity difference $\Delta\rho$ from that of end members BaAl$_4$ and EuAl$_4$, respectively (see main text and Equation~\ref{Eq: Delta Rho}),
   and CDW transition temperature data obtained from resistivity, neutron and x-ray diffraction data. The solid and dashed lines are guides to the eye.}
    \label{fig:PhaseDiagram}
\end{figure*}

We first discuss the effect of isovalent cation substitutions on the crystallographic structure of BaAl$_4$. As shown in Fig.~\ref{fig: X-ray}, there is very minimal change in lattice parameters across both Sr- and Eu-substitution  series. While the $c/a$ ratio increases as expected due to the difference in cation radii, the unit cell only changes by $\sim 1\%$ along the crystallographic $c$-axis across both Sr- and Eu-based series, which is a minute difference when compared to other similar Ba-Sr substitution series such as Ba$_{1-x}$Sr$_x$Fe$_2$As$_2$  \cite{Kirshenbaum-FeAs} and Ba$_{1-x}$Sr$_x$Ni$_2$As$_2$ \cite{Eckberg-Nematic}. 
Geometric optimization calculations of the Ba$_{1-x}$Sr$_x$Al$_4$ experimental structures (performed to achieve less than 0.001~eV/\AA~force on each atom) yield relaxed structures that differ from the corresponding experimental $a$- and $c$-axis lattice constants by a maximum of 0.25\% and 0.50\%, respectively. Our calculations confirm the experimental findings that the crystal structures, e.g., $c/a$ ratio, remain mostly unchanged with Ba-Sr substitution, suggesting that the Al-substructure is quite rigid in this series and does not permit the usual chemical pressure effect of isovalent alkali earth substitution, such as found in e.g. Ba$_{1-x}$Sr$_x$Fe$_2$As$_2$, or between EuFe$_2$As$_2$ and the Sr and Ba counterparts, where a much larger $\sim$10\% reduction in $c$-axis unit cell dimension is achieved \cite{Kirshenbaum-FeAs,EuAl4-Parameters}. 
This observation is one of the main motivations to try and understand the mechanism behind the strong tuning of CDW order with chemical substitution in the absence of large effects on crystallographic structure.
Interestingly, the change in tetrahedral bond angles $\alpha$ and $\beta$ evolve as expected for chemical pressure being applied to the tetragonal unit cell with Sr substitution, but with a much larger change than the Fe- and Ni-based examples which only vary by $<1 \degree$ from $x$=0 to 1. 
This again reflects the dominance of the Al-substructure.

The metallic resistivity $\rho(T)$ of BaAl$_4$ and the related Sr- and Eu-based members is very similar in the paramagnetic, non-ordered regimes. SrAl$_4$ exhibits metallic behavior on cooling, and the CDW transition $T_{cdw}$ ($x$=1), which appears as a kink in $\rho(T)$ at 235~K, can be tracked as a function of Ba substitution in Ba$_{1-x}$Sr$_x$Al$_4$ as shown in Fig.~\ref{fig:PhaseDiagram}a). Similarly, in Fig. ~\ref{fig:PhaseDiagram}c), EuAl$_4$ exhibits metallic behavior upon cooling, followed by two distinct transitions: a CDW transition at 140 K and an antiferromagnetic transition at 15 K. The transitions in the Eu and Sr compounds are progressively suppressed with increasing Ba substitution, as evidenced by the shift in the temperature of the kink observed in $\rho(T)$. The alkali earth substitution does not otherwise change the overall temperature behavior of $\rho(T)$, with the high temperature linear slope nearly unchanged. 
The effect of substitution and the evolution of $T_{cdw}$ can be clearly demonstrated by plotting the change in $\rho(T)$ as a function of $x$ from the metallic parent compound BaAl$_4$, defined as
\begin{equation}
    \Delta\rho = \frac{\rho_x(T) - \rho_x (2~\rm{K})}{\rho_x (300~\rm{K}) - \rho_{\it{x}} (2~\rm{K}) } - \frac{\rho_{0}(T) - \rho_{0} (2~\rm{K})}{\rho_{0} (300~\rm{K}) - \rho_{0} (2~\rm{K}) }.
    \label{Eq: Delta Rho}
\end{equation}
In this equation, $\rho_x$ is the resistivity of the substituted sample and $\rho_{0}$ is the resistivity of BaAl$_4$. This normalizes the value of resistivity to lie between 0 and 1 and shows the magnitude of the deviation from the metallic resistivity of BaAl$_4$, where $T_{cdw}$=0.
The resultant phase diagram for Ba$_{1-x}$Sr$_x$Al$_4$ is presented in Fig.~\ref{fig:PhaseDiagram}b), which reflects the lack of change in $\rho(T)$ as a function of $x$. For Ba$_{1-y}$Eu$_y$Al$_4$, a similar trend is shown in Fig.~\ref{fig:PhaseDiagram}d), but with weaker deviations from the normalizing curve than for  Ba$_{1-x}$Sr$_x$Al$_4$,
Aside from the onset and evolution of CDW order, which will be discussed below, the near insensitivity of resistivity behavior to isovalent alkali earth substitution is not surprising given the similar electronic band structures \cite{NAKAMURA-SrAl4,Nakamure-EuGa4,BaAl4-Topology,BaAl4-Topology2} in each end member and the dominant aluminum bands \cite{Wang2024,Ramakrishnan}.

\begin{figure*}
\includegraphics[width=0.98\linewidth]{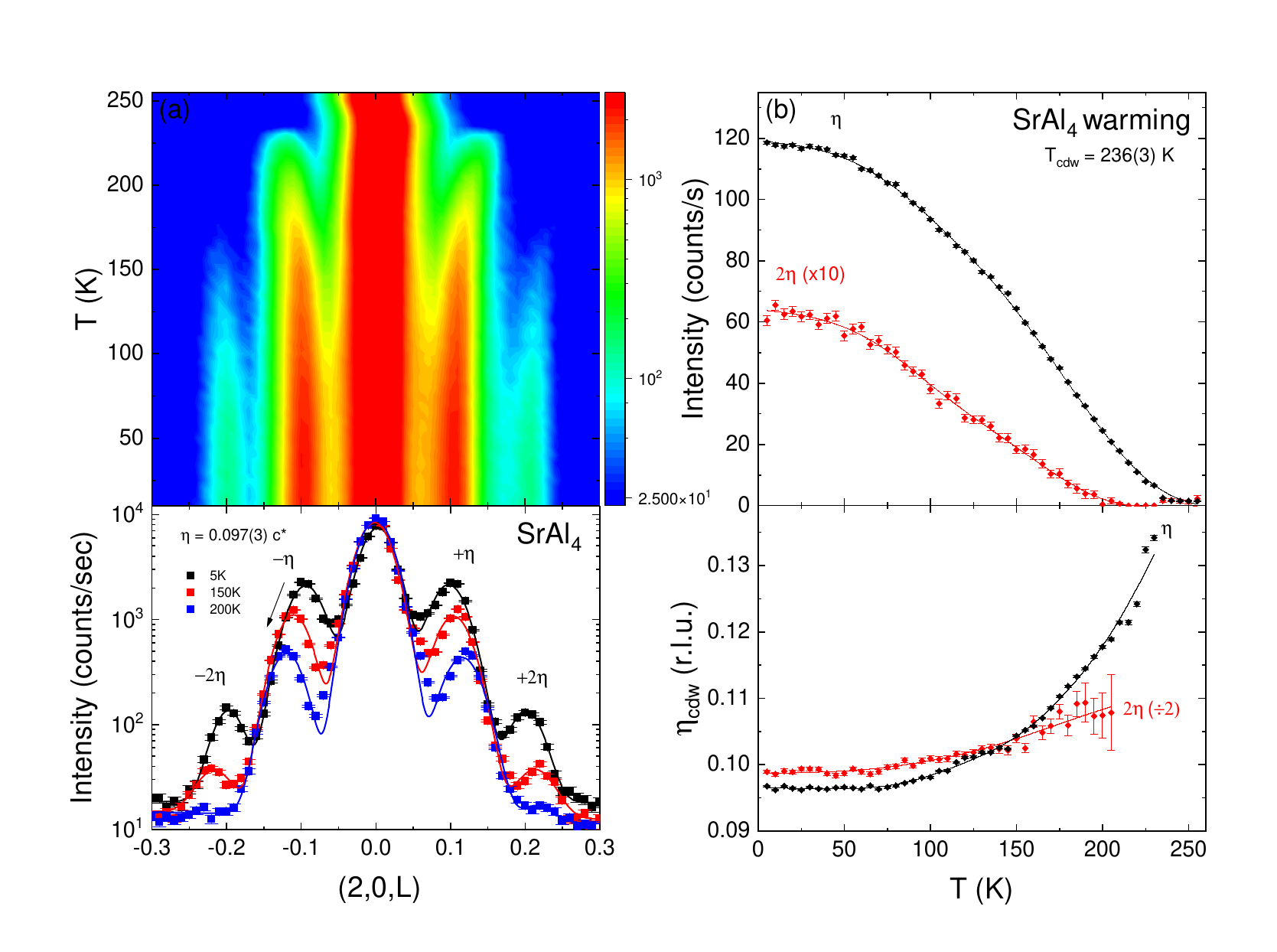}
\caption{Neutron diffraction along (2,0,L) as a function of temperature for SrAl$_4$. (a) The top plot shows a contour map of the scattering intensity showing first- and second- order satellites around the fundamental Bragg peak. The lower plot shows scans at 5, 150 and 200~K on a log scale with Gaussian fits to yield $q_{cdw}$ = $\eta$c$^*$, with $\eta$ = 0.097(3) in the ground state. (b) The top plots show the integrated intensity vs. temperature extracted from Gaussian fits to the data. The  black points represent the intensity of the primary (first order) CDW order parameter and red the second order peak. A mean field  fit to the primary peaks gives a transition temperature of $T_{cdw}$ = 236(3)~K, while the second order peaks become unobservable above $\sim$200~K. The bottom plot shows the temperature dependence of the first and second order wavevectors in reciprocal lattice units extracted from Gaussian fits.  In principle the data should overlap, with the small offset likely due to the overlapping peaks in the fits.  The incommensurate wavevector increases smoothly and continuously with temperature up to $T_{cdw}$, with no evidence of any lock-in transitions.}
\label{fig:Neutron SrAl4}
\end{figure*}

Despite such a small variation in lattice parameters, there is a strong evolution of CDW order with both Sr and Eu substitution, as shown in the phase diagrams of Fig.~\ref{fig:PhaseDiagram}. 
The decrease in $T_{cdw}$ with Ba substitution in both series is akin to the decrease observed as a function of pressure \cite{NAKAMURA-SrAl4,Zhang-Pressure}. However, given the notable difference in crystallographic response noted above for alkali earth substitution vs. applied pressure, it is not clear the mechanism is the same. Furthermore, in the substitution series, $T_{cdw}$ appears to decrease over a wider temperature range before an abrupt drop near the midway point in substitution in both systems:
Ba$_{1-x}$Sr$_x$Al$_4$ appears to show a discontinuous jump in $T_{cdw}$ from zero to $\sim 100$~K near $x$=0.5, and Ba$_{1-y}$Eu$_y$Al$_4$ also exhibits a jump close to $y$=0.5.
Lacking any abrupt changes in crystallographic parameters (c.f. Fig.~1) in either system, this suggests the CDW phase itself approaches a critical point of instability. We will discuss this below.

To characterize the CDW distortion in these series, we performed x-ray and neutron diffraction measurements in the $(h,0,l)$ scattering plane for a series of substitutions through these series. Fig. \ref{fig:Neutron SrAl4}(a) shows neutron diffraction scans for SrAl$_4$ around (2,0,0) measured along $c^*$ = 2$\pi$/c \cite{Lynn-Neutron}. At 5~K, large first-order satellites appear with approximately 30$\%$ of the intensity at the fundamental (2,0,0) Bragg peak. Scans in the $a^*$ direction at $q_{cdw}$ and 2$q_{cdw}$ (not shown) demonstrate that the CDW peaks are strictly along $c^*$. In the ground state of SrAl$_4$, the modulation can be described with an incommensurate wavevector of the form $q_{cdw}$ = (0,0, $\eta$) with $\eta$ = 0.097(3). The smoothly increasing temperature dependence makes the incommensurability clear. A second-order satellite indexed as 2$q_{cdw}$ can be seen more clearly in the Fig. \ref{fig:Neutron SrAl4}(a) bottom plot, which shows the same scan condition at several temperatures. Near 200~K, the intensity at 2$q_{cdw}$ becomes unobservable above the background. The top plot of Fig. \ref{fig:Neutron SrAl4}(a) displays the intensity map of scans along (2,0,L), measured on warming for temperatures up to 255 K. The smooth evolution of the superlattice peaks away from (2,0,0) up to $\sim$ 235 K confirms the incommensurability of the lattice distortion.

\begin{figure*}
\includegraphics[width=\linewidth]{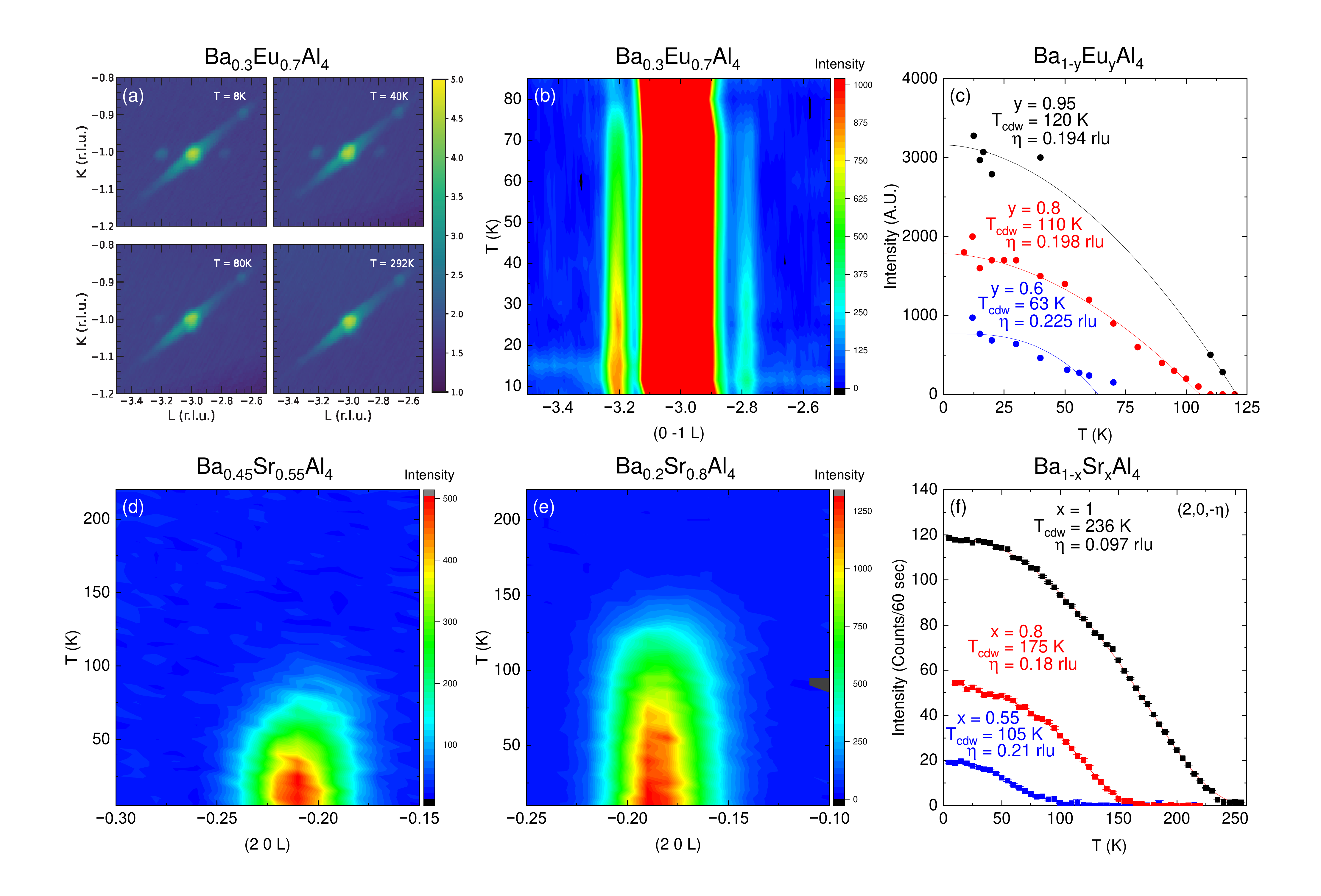}
    \caption{(a) K, L momentum space cut of x-ray data for Ba$_{0.7}$Eu$_{0.3}$Al$_4$ at different temperatures. (b) Contour map of scattering intensity showing satelite peaks around the Bragg peak in Ba$_{0.7}$Eu$_{0.3}$Al$_4$. (c) Intensity as a function of temperature for Ba$_{1-x}$Eu$_{x}$Al$_4$, where x = 0.95, 0.8 and 0.6. (d)-(f) Temperature dependence of the first order CDW superlattice peaks for Ba$_{1-x}$Sr$_x$Al$_4$ (x=0.55 and 0.8) measured with neutrons in the (H,0,L) scattering plane. (d) and (e) contour plots of diffraction intensity for x=0.8 and x=0.55 respectively. The first order peak shows $\eta$ = 0.18 for x = 0.8 and $\eta$ = 0.21 for x = 0.55. (f) Integrated intensity as a function of temperature extracted from Gaussian fits to the data for Ba$_{1-x}$Sr$_{x}$Al$_4$, where x = 1, 0.8 and 0.55.(d) 
    }
    \label{fig:Substituted Neutron}
\end{figure*}

The CDW intensity for the primary CDW satellite peak, which is proportional to the square of the order parameter, is plotted in Fig.~\ref{fig:Neutron SrAl4}(b).  These data were obtained from the integrated intensities of one of the satellite peaks extracted from the fits. A simple mean field fit to the intensity at $\eta$ versus $T$ estimates an ordering temperature $T_{cdw}$ = 236(3)~K. The intensity of 2$\eta$ falls off more rapidly, vanishing around 200 K. The wavevectors extracted from Gaussian fits of the (2,0,L) scans (bottom of Fig.\ref{fig:Neutron SrAl4}(b)) show a smooth increase up to the phase transition. The results are consistent with a CDW with a single wavevector of the form $\eta q$. We note that for metallic systems, the CDW wavevector is typically controlled by Fermi surface nesting and therefore does not show significant temperature dependence.  This suggests that in the present system either the Fermi surface exhibits a significant variation with temperature, or $q_{cdw}$ is not directly associated with the Fermi surface, rather is controlled by competing interactions \cite{Sangjun-CDW}. We will discuss this below.

\begin{figure}
\includegraphics[width=0.45\textwidth]{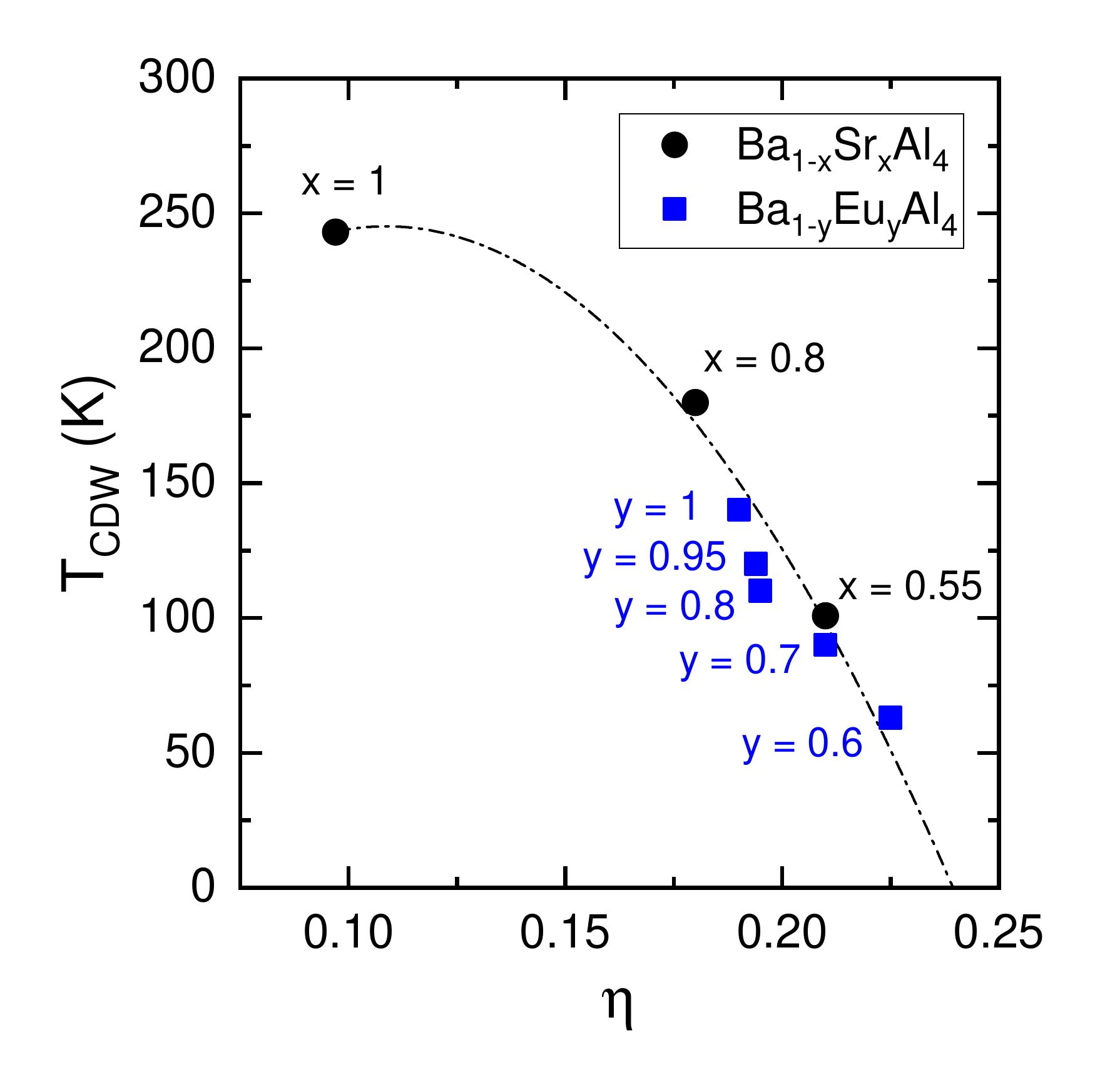}
\caption{T$_{cdw}$ vs. $\eta$ for various members of the $A$Al$_4$ family. Values for EuAl$_4$ have been taken from \cite{Kaneko2021}.}
\label{fig:EtaVTepmerature}
\end{figure}

\begin{figure}[htbp]
  \centering
  \includegraphics[scale=0.3, width=0.45\textwidth]{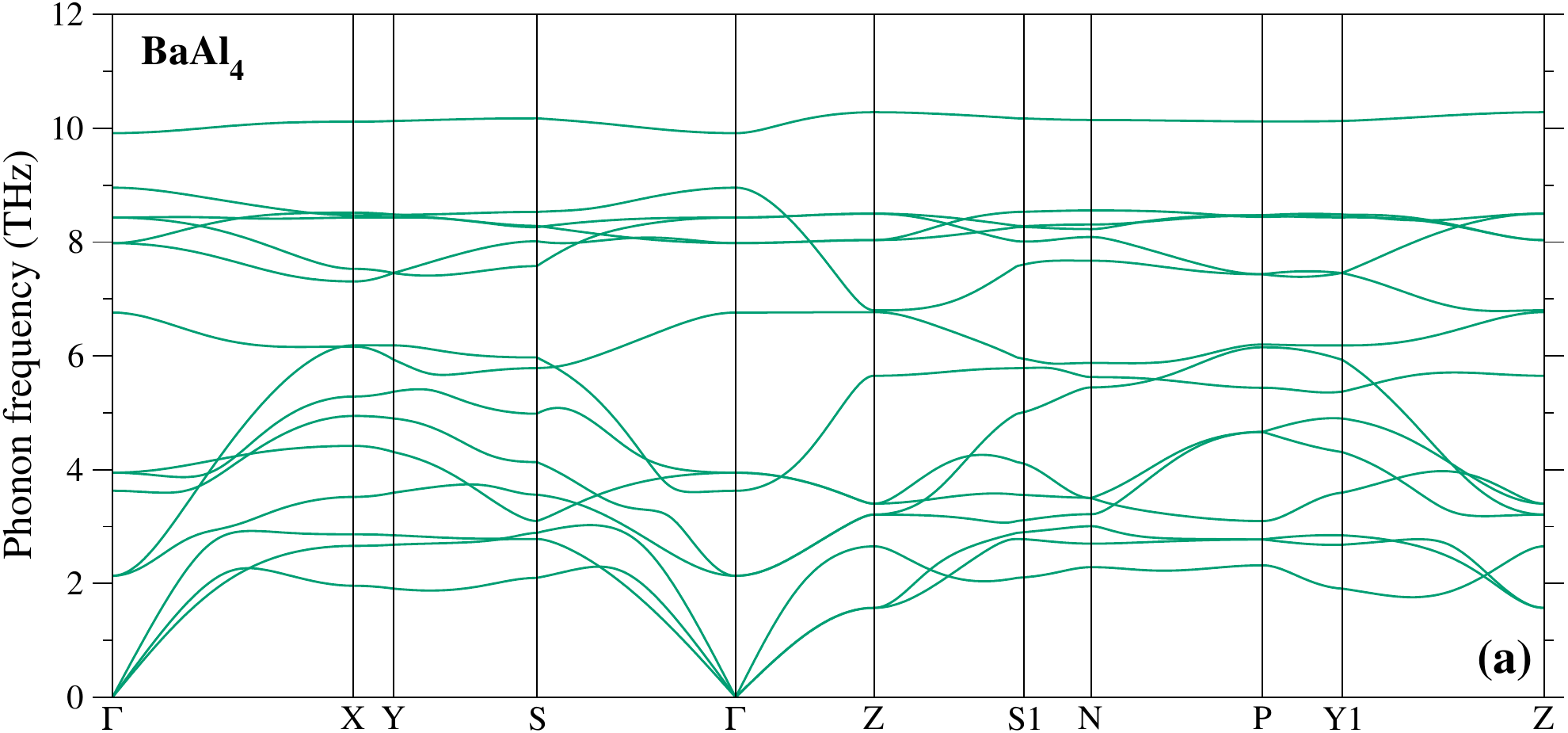} \\
  \vspace{10pt}
  \includegraphics[scale=0.3, width=0.45\textwidth]{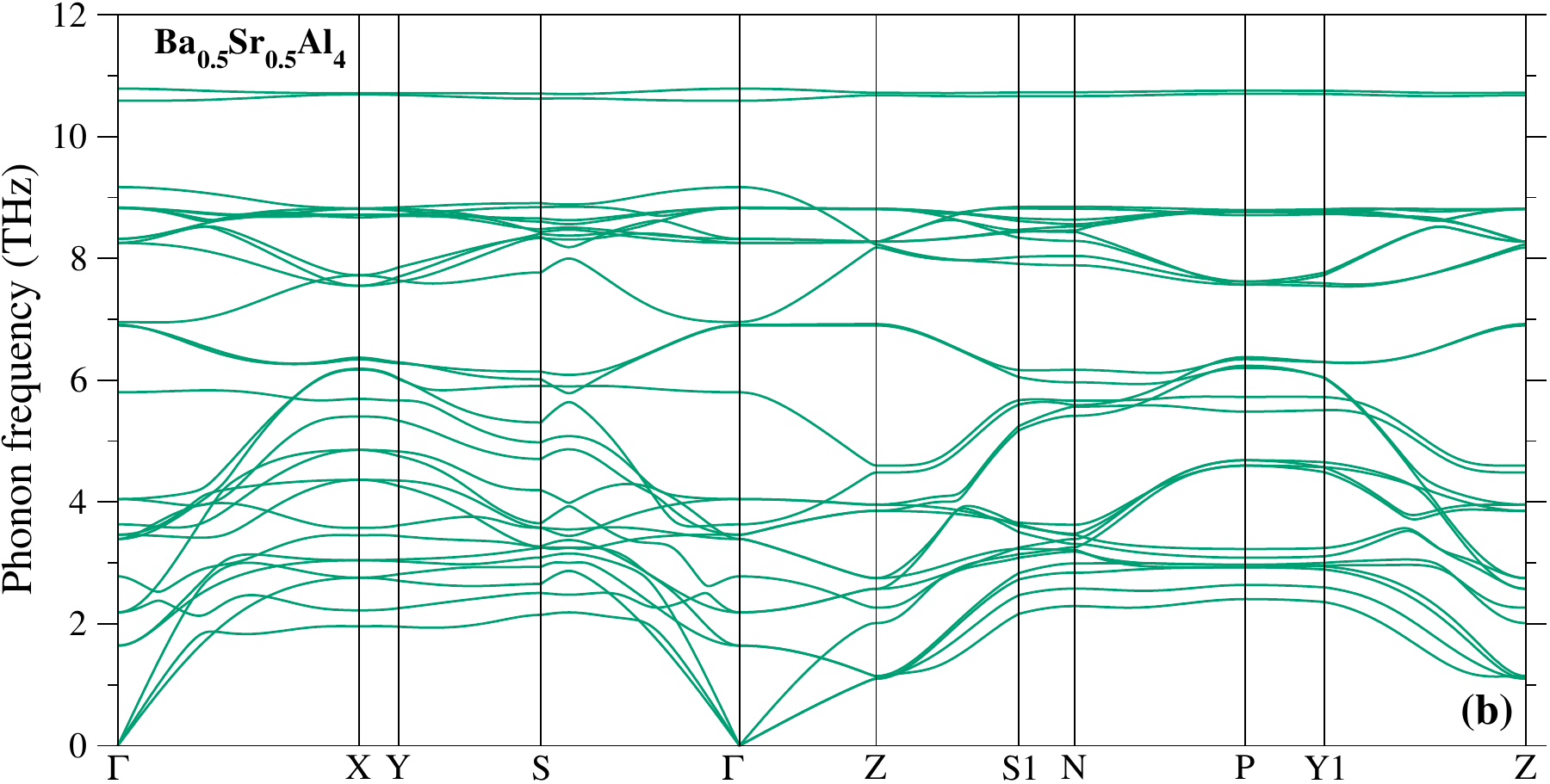} \\
  \vspace{10pt}
  \includegraphics[scale=0.3, width=0.45\textwidth]{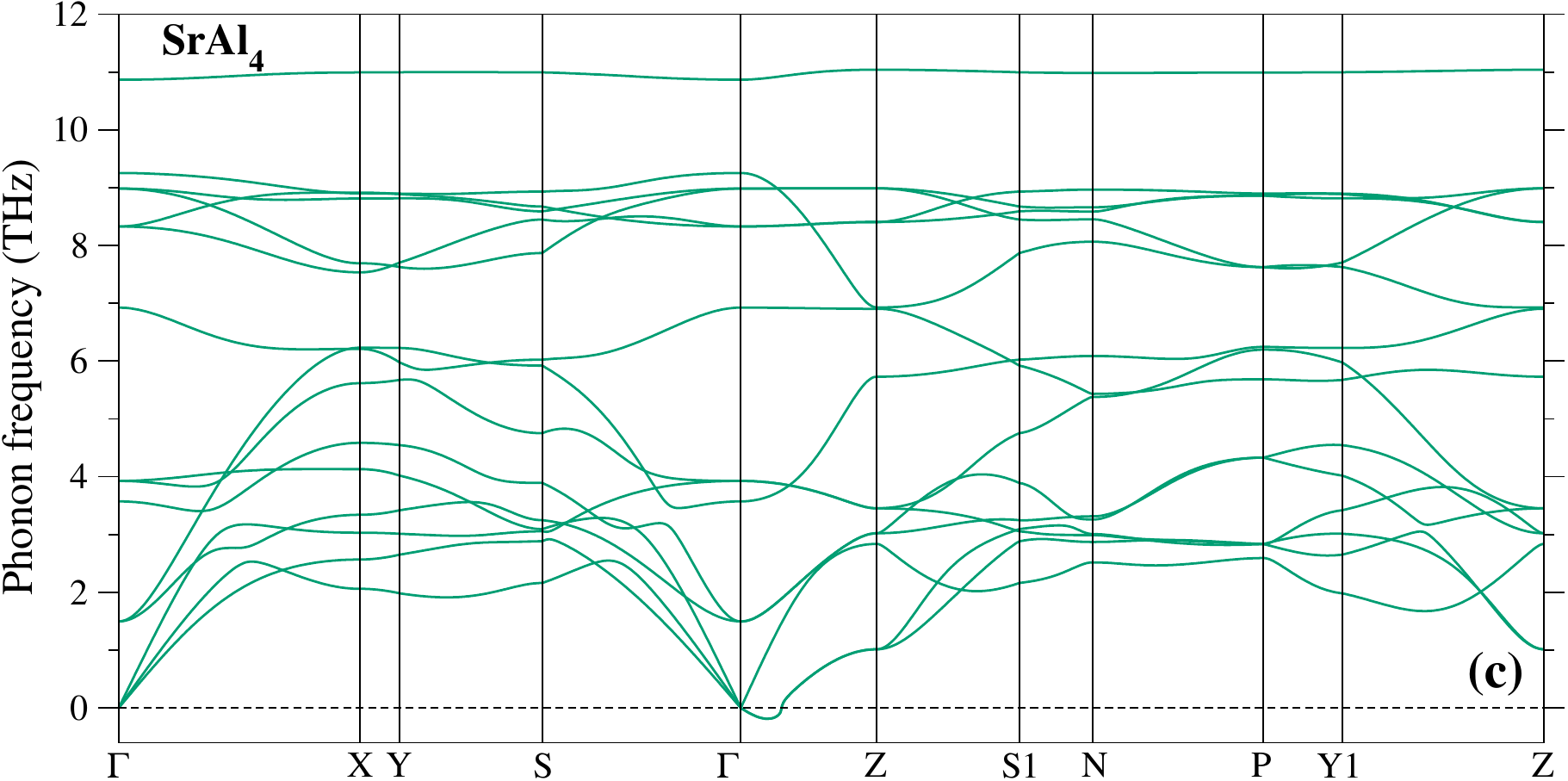}
  \caption{Phonon dispersions for (a) BaAl$_4$, (b) Ba$_{0.5}$Sr$_{0.5}$Al$_4$ and (c) SrAl$_4$ along the Brillouin zone path of the corresponding primitive cells.}
  \label{fig:phonons}
\end{figure}

To investigate the CDW evolution with cation substitutions, we repeated temperature-dependent neutron diffraction measurements for single crystals of  Ba$_{1-x}$Sr$_x$Al$_4$ ($x$=0.55 and 0.80) and X-ray diffraction measurements of Ba$_{1-y}$Eu$_y$Al$_4$ ($y$=0.6, 0.80 and 0.95).
The resultant data for both series are summarized in Fig. \ref{fig:Substituted Neutron}, with integrated intensities plotted as a function of temperature in the same manner as above for the first-order CDW satellites. The CDW transition temperature decreases with larger Ba concentrations in both series down to $T_{cdw}$ = 105(3) K for $x$=0.55, and 63 K for $y$=0.6, both just above the critical 50\% onset concentration. More notably, in both series Ba substitutions causes an increase in the CDW wavevector of the form q$_{cdw}$ = (0,0, $\eta$), with $\eta$ nearly doubling for both series, with progressive increases in $\eta$ as $T_{cdw}$ decreases.
Presented in Fig.~\ref{fig:EtaVTepmerature}, this universal scaling, i.e. $\eta(T_{cdw})$, completely captures the evolution of incommensurability in both Sr and Eu-based series, including the previously reported value $\eta$ = 0.19 for EuAl$_4$ with $T_{cdw}$ = 140 K \cite{Kaneko2021}. 
This suggests the correlation between $T_{cdw}$ and $\eta$ is a direct reflection of the driving mechanism of CDW order in this family.
Note that while $\eta$ remains clearly incommensurate across the series, its amplitude (not shown) does weaken with decreasing ordering temperature. Moreover, the trend in both Sr- and Eu-based series toward  $T_{cdw}$=0 near  $\eta \simeq 0.23$ suggests an explanation for the abrupt drop in $T_{cdw}$=0 as a function of substitution in both systems (i.e., at $x$=0.5 and $y$=0.5, c.f. Fig.~2)

\section{Phonon Band Structure}

A recent computational study \cite{Wang2024} considered the origin of CDW order in the $A$Al$_4$ series, comparing the traditional Peierls-type Fermi surface nesting scenario \cite{Peierls} with a strong electron-phonon coupling scenario. Wang et al. found that the latter interaction leads to softening of the transverse acoustic (TA) mode that becomes imaginary at $q$ = 0.24($\pi$/$c$), reported to be in good agreement with the original cited value \cite{NAKAMURA-SrAl4}.
Here, we calculate the phonon dispersions for $x$= 0, 0.5 and 1, and compare them to Wang et al. and our experimental data. 

DFT calculations were performed to obtain phonon dispersions using the finite difference method for the three structures with experimental lattice parameters (for $x$=0.50, we use  the experimental lattice constants obtained for $x=0.55$). We used the Vienna {\it ab initio} simulation package (VASP)~\cite{vasp} + PHONOPY~\cite{phonopy} implementations to calculate the phonon dispersions, with plane-wave basis set (cutoff of 650 eV) and projected augmented wave (PAW) pseudopotentials with PBE exchange-correlation functional including spin-orbit coupling (SOC). For the parent compounds, we have used a supercell of size $2\times 2\times 2$ of the primitive cell, and the convergence of the phonon bands was checked with respect to the $k$-point sampling ( 14 $\times $ 14 $\times$ 14 ) with Gaussian smearing of width 0.05 eV. For the $x$=0.50 structure, a supercell of size $2\times 2\times 2$ of the conventional cell was used with $k$-point mesh size $10\times 10\times 4$. 

The calculated phonon dispersions are plotted in Fig.~\ref{fig:phonons} for $x$= 0, 0.5 and 1. In agreement with previous studies, we found no phonon instabilities in the parent compound BaAl$_4$. For SrAl$_4$, there is a softening of the TA mode along the $\Gamma$-Z direction, 
with a minimum in the dispersion at 0.095$(2\pi/c)$, consistent with the previous calculation which used a larger supercell \cite{Wang2024} and in excellent agreement with $\eta$=0.097 obtained from neutron diffraction data reported above.
From our calculation for $x$=0.5, there are no unstable phonon modes present along the $\Gamma$-Z direction, although the frequency of the TA phonon mode at the Z-point is lower compared to BaAl$_4$ and nearly similar to that of SrAl$_4$. The $\Gamma$-Z branch also clearly shows a flattening as compared to BaAl$_4$, suggesting that it is very near the onset of a softened mode. This is consistent with our experimental data, where $\eta$ is also zero at $x$=0.5 but becomes finite (i.e. CDW order onsets) just above at $x=0.55$.

Intriguingly, all the calculated phonon spectra show a flat optical phonon band at the highest frequency  mode (9.8-11~THz). This band is doubly degenerate in the parent compounds and the degeneracy gets lifted in the substituted compounds due to the introduction of more than two distinct atomic positions for Al atoms. The flat band corresponds to modes which involve movement of only Al atoms in the $c$-direction. This degeneracy gets lifted for the $x$=0.5 case as the Al-cage contains two types of center atoms, Ba or Sr.

\section{Discussion and Conclusion}

The insensitivity of the crystallographic structure to alkali earth substitution in the $A$Al$_4$ series makes the strong evolution of CDW order surprising in light of a seemingly unchanging chemical bonding environment. Furthermore, the stark similarities between CDW order in both Ba$_{1-x}$Sr$_x$Al$_4$ and Ba$_{1-y}$Eu$_y$Al$_4$ series, in particular the newly observed universal relation between ordering temperature $T_{cdw}$ and ordering vector $\eta$ (c.f. Fig.\ref{fig:EtaVTepmerature})
suggest an underlying mechanism that is rather independent of details of cation bonding or even magnetic interactions.
Previous de Haas van Alphen measurements \cite{NAKAMURA-SrAl4} characterizing the Fermi surfaces of the $A$Al$_4$ series suggested that the Fermi surface of BaAl$_4$ is different from that of SrAl$_4$, with the latter having a closer resemblance to that of its magnetic counterpart EuAl$_4$, leading Nakamura et al. to suggest the existence of a nesting wavevector in SrAl$_4$. However, recent calculations of the Fermi surfaces of BaAl$_4$ \cite{BaAl4-Topology}, SrAl$_4$ and EuAl$_4$ \cite{Wang2024,Ramakrishnan} have found that these are semimetals with nearly identical electronic environments, with strong spin-orbit coupling situating Dirac points in the spectra about 0.2-0.3~eV above the Fermi energy.
Moreover, recent calculations of the band structure of SrAl$_4$ \cite{Wang2024,Ramakrishnan} have found an imperfect nesting, leading Wang et al. to the conclusion that strong electron-phonon coupling, rather than nesting, yields a softened TA phonon mode with small $q$-vector along $\Gamma$-Z, explaining the origin of CDW order in the series \cite{Wang2024}.  
This is in line with our findings, including both the calculated evolution of the TA modes and measured ordering wavevector through the Ba$_{1-x}$Sr$_x$Al$_4$ series, and suggests the variation in the TA phonon mode stem from changes in the cation mass, nicely demonstrated by the continuous diminishing of CDW order with increasing ratio of Ba to Sr and comparisons to EuAl$_4$ \cite{Ramakrishnan,Ramakrishnan-EuAl4}.

However, questions remain about CDW order and its sensitivity to atomic structure and substitution. Calculations by Ramakrishnan et al indeed find an imperfect nesting in the electronic band structure of SrAl$_4$, but in contrast to the results of this study and Wang et al., they did not identify any imaginary or soft modes attributable to the mechanism of CDW formation, concluding that the mechanism remains unknown \cite{Ramakrishnan}.

Furthermore, while it has been known that the $x$=0 and $x$=1 end members of the Ba$_{1-x}$Sr$_x$Al$_4$ series necessitate a transition between CDW and non-ordered ground states at an intermediate concentration, 
our study is the first to report an unexpected abrupt cutoff in the CDW phase near 50\% substitution concentrations, observed in both series and consistent with the universal relation between ordering vector and temperature.
Given the assumed smooth variation of the TA phonon mode with changing cation mass is the driving force, an abrupt jump in the spectrum is not necessarily expected, suggesting either a percolation threshold is reached or another energy scale plays an intermediate role in the destabilization of CDW order. Considering both Sr- and Eu-based series show this instability at the same critical c-axis ordering vector, the approach to this length scale warrants further investigation.

There are a few systems isostructural to SrAl$_4$ that harbor charge order and competing interactions. In particular, recent studies of the evolution of structural and charge order in the Ba$_{1-x}$Sr$_x$Ni$_2$As$_2$ \cite{CDW-BaNi2As2,Eckberg-Nematic,Dushyant-BaNi2As2} have identified charge order phases with similar temperature scales as SrAl$_4$. While these systems share an incommensurate nature, the NiAs-based system exhibits multiple coexisting CDW phases with much shorter wavelengths, and more important, strong correlations 
between structural distortions, CDW order and electronic nematic fluctuations and ordered phases. Interestingly, the Ba$_{1-x}$Sr$_x$Ni$_2$As$_2$ phase diagram exhibits an abrupt drop of its CDW order near 70\% Sr not unlike the discontinuous change near 50\% in both Ba$_{1-x}$Sr$_x$Al$_4$ and Ba$_{1-y}$Eu$_y$Al$_4$, suggesting competing phases or intermediate energy scales may be playing a role in both cases.
Regardless of the underlying mechanism of the first-order suppression of CDW order in the $A$Al$_4$ series reported here, it is interesting to consider the utility of this region of the phase diagram for tuning the electronic properties. Given the proven non-trivial topology of the electronic structures of both end members, this presents a unique ability to switch between the presence of CDW order while maintaining topological band structure, suggesting a new route to controlling the coupling of electronic interactions with band structure topology in a relatively inert and simple system.

Finally, a lower temperature first-order transition near 90~K was previously observed in the electrical resistivity of SrAl$_4$ \cite{NAKAMURA-SrAl4}, but is not observed in our study of Ba$_{1-x}$Sr$_x$Al$_4$ single-crystal samples. This transition was recently reported to be a structural distortion from the high temperature tetragonal phase to a monoclinic phase \cite{Ramakrishnan}. Because it was also reported that this distortion leaves the CDW order virtually unaffected, we do not believe it has significant consequences for our interpretations. 

In this work we have studied the structural and physical properties of the Ba$_{1-x}$Sr$_x$Al$_4$ and Ba$_{1-y}$Eu$_y$Al$_4$ solid solution series to investigate the evolution of charge density wave order with cation substitution up to its termination near 50\% substitution in both series. Using neutron and X-ray diffraction measurements, we have determined the wavevectors of $c$-axis-oriented charge order to evolve smoothly in both series, ranging from $\eta$ = 0.097 in SrAl$_4$ to $\eta$ 0.225 in Ba$_{0.4}$Eu$_{0.6}$Al$_4$, with a common trend in both systems that depends intrinsically on the ordering temperature and is independent of the type of cation substitution.
A comparison to phonon band calculations in Ba$_{1-x}$Sr$_x$Al$_4$ confirms a mechanism for CDW order stemming from softening of a transverse acoustic phonon mode with small wavevector along the $\Gamma$-$Z$ direction, consistent with prior results \cite{Wang2024}. 
This work provides a systematic insight into methods of controlling the coupling of electronic interactions with band structure topology in a simple yet versatile platform.

\section{Acknowledgments}

Research at the University of Maryland was supported by the Gordon and Betty Moore Foundation’s EPiQS Initiative through Grant No. GBMF9071, the U.S. National Science Foundation Grant No. DMR2303090, the Binational Science Foundation Grant No. 2022126, and the Maryland Quantum Materials Center.
S.B. and R.V. acknowledge support by the Deutsche Forschungsgemeinschaft (DFG, German Research Foundation) for funding through project TRR 288 — 422213477 (project A05). 
S.B. also acknowledges support from the US National Science Foundation (NSF) Grant Number 2201516 under the Accelnet program of Office of International Science and Engineering (OISE). X-ray experiments were supported by the Center for Quantum Sensing and Quantum Materials under DOE BES Award No. DE-SC0021238. P.A. acknowledges support from the EPiQS program of the Gordon and Betty Moore Foundation, Grant No. GBMF9452. The identification of any commercial product or trade name does not imply endorsement or recommendation by the National Institute of Standards and Technology

\bibliography{BaSrAl4Refs_resub1}

\end{document}